\documentclass[twocolumn,times]{aastex62}
\usepackage{graphicx}
\usepackage[flushleft]{threeparttable}
\usepackage{blindtext}
\usepackage{amsmath}
\usepackage{mathtools}
\usepackage{multirow}
\turnoffeditone

\newcommand{\lya}{Ly$\alpha$ }
\begin{document}
\shortauthors{Jung et al.}
\def\nar{New Astron.}
\def\na{New Astron.}
\title{\large \textbf{Texas Spectroscopic Search for \lya Emission at the End of Reionization \\II. The Deepest Near-Infrared Spectroscopic Observation at $z \gtrsim 7$}}

\submitjournal{\textit{the Astrophysical Journal}}

\correspondingauthor{Intae Jung}
\email{itjung@astro.as.utexas.edu}

\author[0000-0003-1187-4240]{Intae Jung}
\affil{Department of Astronomy, The University of Texas at Austin, Austin, TX 78712, USA}

\author[0000-0001-8519-1130]{Steven L. Finkelstein}
\affil{Department of Astronomy, The University of Texas at Austin, Austin, TX 78712, USA}

\author[0000-0001-5414-5131]{Mark Dickinson}
\affil{National Optical Astronomy Observatory, Tucson, AZ 85719, USA}

\author[0000-0001-6251-4988]{Taylor A. Hutchison}
\affil{Department of Physics and Astronomy, Texas A\&M University, College
Station, TX, 77843-4242 USA}
\affil{George P.\ and Cynthia Woods Mitchell Institute for
  Fundamental Physics and Astronomy, Texas A\&M University, College
  Station, TX, 77843-4242 USA}
  
\author[0000-0003-2366-8858]{Rebecca L. Larson}
\affil{Department of Astronomy, The University of Texas at Austin, Austin, TX 78712, USA}

\author[0000-0001-7503-8482]{Casey Papovich}
\affil{Department of Physics and Astronomy, Texas A\&M University, College
Station, TX, 77843-4242 USA}
\affil{George P.\ and Cynthia Woods Mitchell Institute for
  Fundamental Physics and Astronomy, Texas A\&M University, College
  Station, TX, 77843-4242 USA}

\author[0000-0001-8940-6768]{Laura Pentericci}
\affil{INAF, Osservatorio Astronomico di Roma, via Frascati 33, 00078, Monteporzio Catone, Italy}

\author[0000-0002-8442-3128]{Mimi Song}
\affil{Astrophysics Science Division, Goddard Space Flight Center, Greenbelt, MD 20771, USA}

\author[0000-0001-7113-2738]{Henry C. Ferguson}
\affil{Space Telescope Science Institute, 3700 San Martin Drive, Baltimore, MD 21218, USA}

\author[0000-0003-2775-2002]{Yicheng Guo}
\affil{Department of Physics and Astronomy, University of Missouri, Columbia, MO, USA}

\author[0000-0002-9226-5350]{Sangeeta Malhotra}
\affil{Astrophysics Science Division, Goddard Space Flight Center, Greenbelt, MD 20771, USA}
\affil{School of Earth \& Space Exploration, Arizona State University, Tempe, AZ 85287, USA}

\author{Bahram Mobasher}
\affil{Department of Physics and Astronomy, University of California, Riverside, CA 92521, USA}

\author[0000-0002-1501-454X]{James Rhoads}
\affil{Astrophysics Science Division, Goddard Space Flight Center, Greenbelt, MD 20771, USA}
\affil{School of Earth \& Space Exploration, Arizona State University, Tempe, AZ 85287, USA}

\author[0000-0001-8514-7105]{Vithal Tilvi}
\affil{School of Earth \& Space Exploration, Arizona State University, Tempe, AZ 85287, USA}

\author[0000-0002-0784-1852]{Isak Wold}
\affil{Astrophysics Science Division, Goddard Space Flight Center, Greenbelt, MD 20771, USA}

\begin{abstract}
Realizing the utility of \lya emission to trace the evolution of the intergalactic medium (IGM) during the epoch of reionization requires deep spectroscopy across the boundary of optical and near-infrared (NIR) spectrographs at $z\sim7.2$ when \lya emission is at $\sim$1$\mu$m.  Our Texas Spectroscopic Search for \lya Emission at the End of Reionization includes 18 nights of deep spectroscopic observations using the Keck DEIMOS (optical) and MOSFIRE (NIR) spectrographs.  Within this dataset we observe \lya emission from 183 photometric-redshift selected galaxies at $z =$ 5.5 -- 8.3 from the Cosmic Assembly Near infrared Deep Extragalactic Legacy Survey (CANDELS). Our overlapping MOSFIRE observations, over 84 galaxies total, provide the deepest NIR spectroscopic data yet obtained for \lya from galaxies $z > 7$, with $>16$ hr integration time for \textit{four} observed galaxies. Here we analyze these four targets, and we report the discovery of a new $z = 7.60$ \lya detection as well as provide an updated observation of the previously confirmed $z=7.51$ \lya emission from Finkelstein et al. (2013) with a $\sim$3$\times$ longer exposure time. Our analysis of these \lya emission line profiles reveal a significant asymmetric shape.  The two detected \lya emission lines from bright sources ($M_{\text{UV}}<-20.25$) could imply that these bright galaxies inhabit ionized bubbles in a partially neutral IGM, although deeper exposures may yet reveal \lya emission in the fainter sources.
\end{abstract}


\section{Introduction}
Charting the timeline of reionization through useful tracers such as \lya forest absorption in high-$z$ quasars \citep[e.g.,][]{Becker2001a, Fan2006a, Bolton2011a, Mortlock2011a, McGreer2015a, Bosman2018a}, the cosmic microwave background (CMB) polarization measurement \citep{Larson2011a,Planck2016a} and \lya emitter (LAE) observations \citep[e.g.,][]{Miralda1998a, Rhoads2001a, Malhotra2004a}, constrains how galaxies and the intergalactic medium (IGM) interplay in the early universe.   As the dominant sources of the ionizing photons are thought to be galaxies \citep[e.g.,][]{Finkelstein2015a, Robertson2015a}, investigating the evolution of the IGM during reionization provides critical constraints on the evolution of distant galaxies in the early universe as well as the impact of the IGM on the formation and evolution of galaxies at that epoch.

\lya emission has emerged as a useful tracer of the evolution of the IGM near the end of reionization \citep[e.g.,][]{Becker2018a}, as \lya emission is easily diminished with even small amount of neutral hydrogen due to the resonant nature of \lya scattering with neutral hydrogen \citep[e.g.,][]{Rybicki1999a, Santos2004a, Dijkstra2014a}. For instance, narrow-band \lya surveys provide a statistical number of Lyman-alpha emitters (LAEs) for \lya luminosity functions (LFs), and the evolution of the \lya LF at $z\gtrsim6$ suggests an increasing fraction of neutral hydrogen in the IGM \citep[e.g.,][]{Ouchi2010a, Hu2010a, Kashikawa2011a, Zheng2017a, Konno2018a}.  From follow-up spectroscopic observations for high-$z$ candidate galaxies or Lyman-break galaxies (LBGs) a simple measure of the \lya fraction, which is the number of \lya emitters among the number of spectroscopically observed candidates, shows an apparent deficit of \lya emission at $z > 6.5$.  The drop in \lya emission at $z>6$ implies that the HI fraction in the IGM increases significantly from $z\sim6$ $\rightarrow$ 7 \citep[e.g.,][]{Stark2010a, Fontana2010a, Pentericci2011a, Pentericci2014a, Curtis-Lake2012a, Mallery2012a, Caruana2012a, Caruana2014a, Ono2012a, Schenker2012a, Schenker2014a, Treu2012a, Treu2013a, Tilvi2014a, Vanzella2014a, Schmidt2016a}. 

Recently, with extensive \lya spectroscopic data of $\gtrsim60$ \lya detected galaxies at $z\sim6$ -- 7, \cite{Pentericci2018a} suggests a smoother evolution of the IGM with their measurement of the \lya fraction at $z\sim6$ -- 7, where they find little evolution from $z\sim5 \rightarrow 6$ and a larger drop from $z\sim6 \rightarrow 7$.  This reveals that the IGM was not fully ionized by $z =$ 6, thus a smaller evolution in the neutral fraction from $z =$ 6 to 7 is needed to explain the observations.  An analogous analysis of the \lya fraction becomes very challenging at $z>7$. Although spectroscopic follow-up observations with ground-based telescopes and the {\it Hubble Space Telescope} (\textit{HST}) grism have been successful in searching for \lya emission at $z\sim7$ \citep[e.g.,][]{Fontana2010a, Shibuya2012a, Pentericci2014a, Larson2018a}, only six \lya emitting galaxies have been detected so far at $z>7.5$ \citep{Finkelstein2013a, Oesch2015a, Zitrin2015a, Song2016b, Laporte2017a, Hoag2018a}. These non-detections may imply a further drop in the IGM neutral fraction, but this interpretation is non-trivial given the limited spectroscopic depths of most previous NIR spectroscopic observations, and the uncertainty in the expected line wavelength due to the uncertainty of photometric redshift measurements.

In our first paper in this series \citep{Jung2018a} from our \textit{Texas Spectroscopic Search for \lya Emission at the End of Reionization}, we introduced our methodology for constraining the evolution of the \lya EW distribution accounting for all observational incompleteness effects (e.g., photometric redshift probability distribution function, UV continuum luminosity, instrumental wavelength coverage, and observing depth).  We found evidence that the \lya EW distribution evolves to lower values at $z>6$, suggesting an increasing neutral hydrogen fraction in the IGM. To move to $z>7$ we require NIR spectroscopy.  

We obtained deep NIR spectroscopic data with Keck/MOSFIRE over 84 candidate galaxies.  Because these observations partially overlapped on the sky, we achieved $\gtrsim$16hr integration time for four high-$z$ candidate galaxies at $z\gtrsim7$. In this paper, we present the results from these ultra-deep NIR spectroscopic observations with MOSFIRE for four $z\gtrsim7$ galaxies, reporting a new \lya emission line at $z=7.60$ as well as the updated measurement of the previously reported $z=7.51$ \lya emitter \citep{Finkelstein2013a} with a $\sim$3$\times$ longer exposure time. We describe our MOSFIRE datset and data reduction procedures in Section 2, and report the detected \lya emission lines at $z>7$ in Section 3. Section 4 summarizes our findings with our deepest NIR observations and discusses the \lya visibility.  In this work, we assume the $Planck$ cosmology \citep{Planck2016a} with $H_0$ = 67.8\,km\,s$^{-1}$\,Mpc$^{-1}$, $\Omega_{\text{M}}$ = 0.308 and $\Omega_{\Lambda}$ = 0.692.  The \textit{Hubble Space Telescope (HST)} F435W, F606W, F775W, F814W, F850LP, F105W, F125W, F140W and F160W bands are referred as $B_{435}$, $V_{606}$, $i_{775}$, $I_{814}$, $z_{850}$, $Y_{105}$, $J_{125}$, $JH_{140}$ and $H_{160}$, respectively.  All magnitudes are given in the AB system \citep{Oke1983a}, and all errors presented in this paper represent 1$\sigma$ uncertainties (or central 68\% confidence ranges), unless stated otherwise.

\begin{table*}[]
\centering
\caption{Summary of Keck/MOSFIRE observations in GOODS-N}
\begin{tabular}{cccccccc}
\tableline 
\tableline
\quad {Mask Name} & {R.A.}         & {Decl.}         & {Observation Dates} & {$N_{\text{targets}}$} & {$t_{\text{exp}}$} & {Seeing\tablenotemark{a}} &{Standard Star\tablenotemark{b}}\\
\quad {}     & {(J2000.0)} & {(J2000.0)} & {} &{}& {(hrs)} &{(arcsec)}& {}\\
\tableline
\quad {GOODSN\_Mask1} & {189.162917} & {62.274244} & {2013 Apr18}	 	&{24}	& {5.8} &{0.7} &{HIP56147}\\
\quad {GOODSN\_Mask2} & {189.312875} & {62.279597} & {2013 Apr19} 	 	&{19}	& {5.5} &{0.6} &{HIP56147}\\
\quad {GOODSN\_Y\_v12} & {189.244875} & {62.274253} & {2014 Mar14, 15, 25} &{23}	& {6.3} &{0.9} &{HIP53735, HIP65280}\\
\quad {gdn1404\_Y1\_3} & {189.339667} & {62.324689} & {2014 Apr17, 18, May13} &{13}	& {7.2} &{1.3} &{HIP65280}\\
\quad {Mask2\_Y\_2015A} & {189.214083} & {62.265297} & {2015 Feb 23, 24} 		&{10}	& {4.5} &{0.8} &{HIP56147}\\
\quad {Mask1\_Y\_2015A\_2} & {189.331125} & {62.204139} & {2015 Feb 23, 24} 	&{10}	& {4.5} &{1.2} &{HIP56147}\\
\tableline
\end{tabular}
\begin{flushleft}
\tablenotetext{a}{Full width at half maximum measured from continuum objects in mask configurations.}
\tablenotetext{b}{The flux calibration standard stars in our long-slit observations, listed in the Hipparcos index \citep{vanLeeuwen2007a}.}
\end{flushleft}
\label{tab:targets}
\end{table*}

\begin{table*}[]
\centering
\begin{center}
\caption{Summary of four targets with $t_{\text{exp}}>16$ hours}
\begin{tabular}{cccccccc}
\tableline 
\tableline
\quad {ID\tablenotemark{a}} & {R.A.}         & {Decl.}         & {$t_{\text{exp}}$} & {$M_{\text{UV}}$\tablenotemark{b}} & {$z_{\text{phot}}$} & {$z_{\text{spec}}$\tablenotemark{c}} & {EW$_{\text{Ly}\alpha}$\tablenotemark{d}}\\
\quad {}     & {(J2000.0)} & {(J2000.0)} & {(hrs)} & {} &{} &{} & {(\AA)}\\
\tableline
\quad{z7\_GND\_18869} & {189.205292} & {62.250767} & {16.5} &{-19.63}& {7.08$^{+0.18}_{-0.16}$} &{-} &{$<$103.07}\\
\quad{z7\_GND\_16863} & {189.333083} & {62.257236} & {16.3} &{-21.24}& {7.23$^{+0.28}_{-0.29}$} &{7.60} &{$61.28\pm5.85$}\\
\quad{z7\_GND\_42912\tablenotemark{e}} & {189.157875} & {62.302372} & {16.5 } &{-21.58}& {7.54$^{+0.19}_{-0.18}$} &{7.51} &{$33.19\pm3.20$}\\
\quad{z8\_GND\_9408} & {189.300125} & {62.280358} & {19.0} &{-18.99}& {7.71$^{+0.47}_{-6.19}$} &{-} &{$<$386.61}\\
\tableline
\end{tabular}
\end{center}
\begin{flushleft}
\tablenotetext{a}{The listed IDs are from \cite{Finkelstein2015a}, encoded with their photometric redshifts and the fields in the CANDELS imaging data.}
\tablenotetext{b}{$M_{\text{UV}}$ is the averaged magnitude at 1500\AA, derived from galaxy SED fitting with stellar population synthesis models.}
\tablenotetext{c}{The $z_{\text{spec}}$ measurement errors are $\lesssim0.001$.}
\tablenotetext{d}{$5\sigma$ upper limits for non-detections.}
\tablenotetext{e}{Known as z8\_GND\_5296 in \cite{Finkelstein2013a}.}
\end{flushleft}
\label{tab:targets}
\end{table*}

\section{Data}
\subsection{Texas Spectroscopic Search for \lya Emission\\ at the End of Reionization}
To search for Ly$\alpha$ emission from galaxies in the reionization era, we performed deep spectroscopic observations of candidate galaxies in the GOODS-S and GOODS-N fields from the Cosmic Assembly Near-infrared Deep Extragalactic Legacy Survey (CANDELS; PI's Faber \& Ferguson). This consists of a total of 18 nights of spectroscopic observations targeting 183 galaxies at $z>5$: for 118 galaxies with Keck/DEIMOS (PI: R. Livermore) and 84 galaxies with Keck/MOSFIRE (PI: S. Finkelstein; the majority coming through the NASA/Keck allocation). The entire program is described in \cite{Jung2018a} where we discuss our measure of the \lya EW distribution at $z\sim6.5$ with DEIMOS. The target galaxies were selected from the \cite{Finkelstein2015a} photometric catalog. The selection criteria for our masks prioritizes galaxy brightness and the photometric redshift probability being within the $Y$-band instrumental wavelength coverage to maximize the chance of detecting \lya emission. In this paper we report a new \lya emission line at $z=7.60$, analyzing our deepest MOSFIRE observations for four $z\gtrsim7$ galaxies. The \lya EW distribution analysis using our entire MOSFIRE dataset will be discussed in our future publication. 

\subsection{MOSFIRE Y-band observations in GOODS-N}
Our GOODS-N MOSFIRE dataset was obtained through 10 nights of observations with six different mask designs from April 2013 to February 2015, targeting 72 galaxies at $z\gtrsim6$. Table 1 summarizes our Keck/MOSFIRE observations in GOODS-N. In this paper we present the results from our deepest MOSFIRE dataset for the four candidate galaxies, which are summarized in Table 2.  We used the $Y$-band filter with a 0\farcs7 slit width and a spectral resolution of $\sim3$\AA\ ($R=3500$), covering \lya over a redshift range of $7.0 < z < 8.2$. In our observations, we take 180 sec exposures in individual frames with an ABAB dither pattern (+1\farcs25, -1\farcs25, +1\farcs25, -1\farcs25), thus the positions in the adjacent frames are separated by 2\farcs5.

We note that our GOODS-N MOSFIRE program included six objects with $>$16hr integration time, which were classified as high-$z$ candidates from \cite{Finkelstein2015a}. However, two of the six candidates now appear to more likely be low-$z$ galaxies from an updated photometric redshift measurement including deep $I_{\text{814}}$ and {\it Spitzer}/IRAC photometry (Finkelstein et al.\ 2019, in prep), which were not included in the \cite{Finkelstein2015a} measurement. We note these two likely low-$z$ galaxies show no significant features in their spectra.  We thus focus this paper on the remaining four targets.  Each of the four targets were observed in three of the MOSFIRE masks, resulting in the longest NIR spectroscopic follow-up observation for \lya at $z\gtrsim7$, with a total exposure of $>$16hr. 

\subsection{Data Reduction}
The data were reduced using Version 2018 of the public MOSFIRE data reduction pipeline (DRP)\footnote{http://keck-datareductionpipelines.github.io/MosfireDRP/}, which provides a sky-subtracted, flat-fielded, and rectified two-dimensional (2D) slit spectrum per object with a wavelength solution using telluric sky emission lines. The reduced 2D spectra have a spectral resolution of 1.09\AA\ pixel$^{-1}$ and a spatial resolution of 0\farcs18 pixel$^{-1}$. However, pre-2017 MOSFIRE observations were subject to a noticeable drift of object spectra in the spatial direction along the slit, thought to be due to differential atmospheric refraction \citep[e.g.,][]{Kriek2015a, Song2016b}. We detected systematic slit drifts (up to $\sim$1pixel hr$^{-1}$) in all observations, with a noticeable dependence on airmass. To correct for this drifting of object spectra, we ran the MOSFIRE DRP on each adjacent pair of images, and measured the relative slit drift by marking the position of the spectrum of a star we put on one of our slits.  Correcting for the derived slit drifts, we combined individual DRP outputs to generate all combined 2D spectra. In the combination step, we rejected any bad pixel or cosmic ray by taking sigma-clipped means, and we measured the best-fit Gaussian peak fluxes of the continuum sources as the weight factors of the DRP outputs. 

Flux calibration and telluric absorption correction was done on individual nights, using long-slit observations of a spectro-photometric standard star (listed in Table 1) and the model stellar spectrum of \cite{Kurucz1993a}. The response curve is obtained by dividing the model stellar spectra (scaled to have the known broadband magnitudes of the standard stars) with the observed spectra from the long-slit observations. In the case where each mask was observed on multiple nights, we combined all 2D spectra, which were individually calibrated, to generate a single 2D spectrum per mask design. We validated our calibration  using the known $Y_{105}$ magnitudes of continuum objects in the mask configurations (the magnitudes are from the updated photometric catalog of Finkelstein et al.\ 2019, in prep). As these objects were observed contemporaneously to our science objects, they were observed under identical conditions, compared to the standard stars which may not have been.  We therefore used these objects to calculate and apply a residual normalization correction to our flux calibration array, as the ratio of the known $Y$-band magnitudes of the slit continuum objects to those from the calibrated spectra, typically up to a $\sim$30 -- 50\% effect.

All four targets with the longest exposure were observed in three different mask configurations. Thus, after combing all individual mask spectra, weighted by median-noise levels, we obtained fully-reduced, all-mask-combined, and flux-calibrated 2D spectra for the targets. To extract one-dimensional (1D) spectra, we performed an optimal extraction \citep{Horne1986a} with a $2\farcs5$ spatial aperture.

\section{Results}
We performed a visual inspection on the reduced 2D and 1D spectra, and found a new \lya emission line at $z=7.60$ from z7\_GND\_16863 (Figure 1a), in addition to the visible previously reported \lya detection at $z=7.51$ from z7\_GND\_42912 (Figure 1b; known as z8\_GND\_5296 in \citealt{Finkelstein2013a}). The other two targets do not have any obvious emission features at the expected slit positions in the 2D spectra.  We also ran the automated line search algorithm from \cite{Larson2018a} which searches for significant emission features in 1D spectra. Using a threshold of $5\sigma$ for this automated search, this algorithm finds only these same two reported lines, both at a S/N of $>$10.

The \lya properties for our detected lines are derived from the best-fit asymmetric Gaussian function obtained by running the IDL \texttt{MPFIT} package \citep{Markwardt2009a}. The errors of the derived quantities are estimated via Monte-Carlo (MC) simulations by modulating the 1D spectrum with the 1D noise.  We perform MC simulations, fitting the asymmetric Gaussian function to the simulated 1D spectra which are fluctuated by a Gaussian random deviate equal to the associated 1D noise, and we take the standard deviations for the derived \lya properties from the MC simulation runs. 

\begin{figure*}[t]
\centering
\gridline{\fig{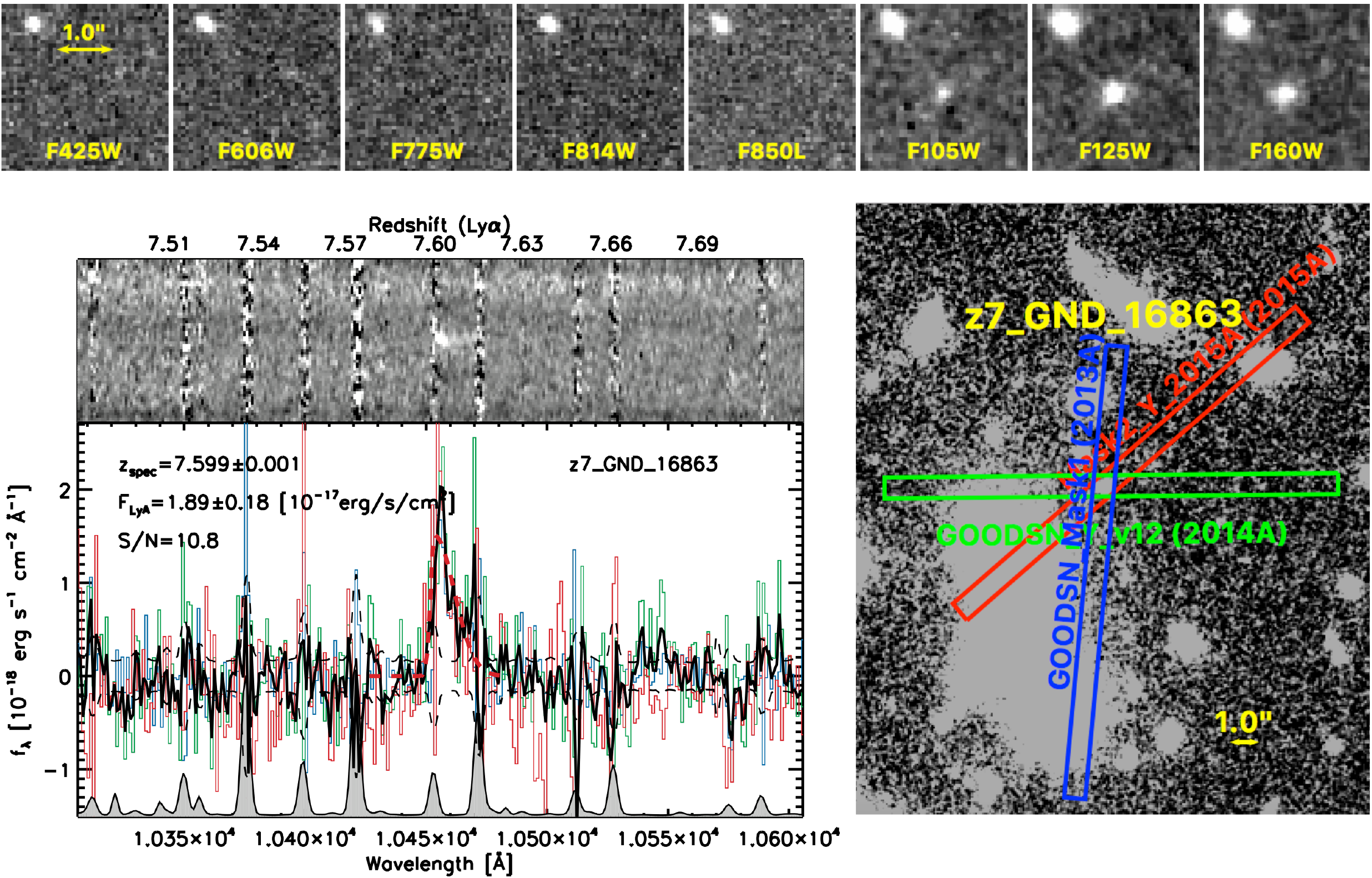}{0.65\paperwidth}{(a) z7\_GND\_16863}}
\gridline{\fig{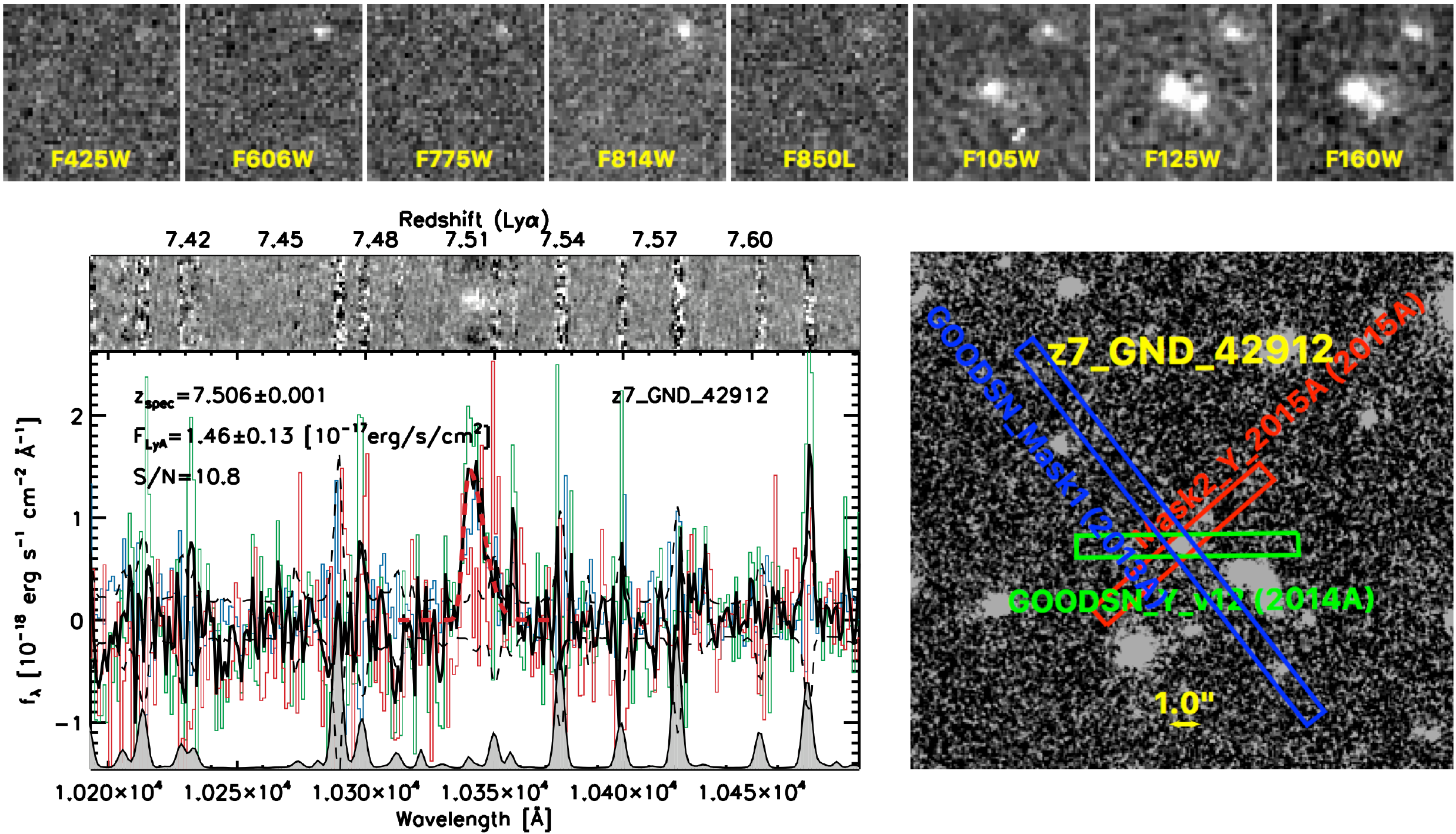}{0.65\paperwidth}{(b) z7\_GND\_42912}}
\caption{(a) The \textit{HST} ACS/WFC3 images of z7\_GND\_16863 in the top panel. All images are centered on the object, and the Lyman break is clearly observed between the \textit{HST} optical and NIR filters. The bottom left panel shows the 1D and 2D spectra of z7\_GND\_16863. In the 2D spectrum, the strong emission line is shown at the expected spatial slit position with the two negative features shown at $\pm2\farcs5$ separation. In the 1D spectrum, the black solid curve is the all-mask-combined flux, and the three individual mask fluxes are displayed as colored histograms.  The dashed black curves represent the $1\sigma$ noise level, and the normalized sky emission is plotted at the bottom as a grey filled curve.  The red dashed curve is the best fit asymmetric Gaussian profile. The bottom right panel shows the slit locations of the three observed mask configurations, overlaid in \textit{HST} WFC3 F160W CANDELS image. The slits are color-coded as histograms in 1D spectra. (b) Same as (a) but for z7\_GND\_42912.}
\label{fig:spectra}
\end{figure*}

\subsection{z7\_GND\_16863: a new \lya detection at $z=7.60$}
As shown in the top row of Figure 1 (a) z7\_GND\_16863 is detected only in the NIR \textit{HST} bands ($Y_{105}$, $J_{125}$, and $H_{160}$), and not in the optical bands ($B_{435}$, $V_{606}$, $i_{775}$, $I_{814}$, and $z_{850}$), suggesting strong continuum break and is consistent with the expectation of a $z=7.60$ galaxy.  This object has been targeted in three mask configurations as shown in the bottom right panel of Figure 1 (a): GOODSN\_Mask2 (2013 Apr), GOODSN\_Y\_v12 (2014 Mar), and Mask2\_Y\_2015A (2015 Feb) with a total exposure time of $t_{\text{exp}}$ = 16.3 hrs. The emission line has been detected at 10450\AA\ ($z_{\text{spec}}=7.60$) with a signal-to-noise (S/N) ratio of 10.8. In the 2D spectrum, the strong emission line is shown at the expected slit position with the two negative features shown at $\pm2\farcs5$ separations. Interestingly, this spectroscopic redshift deviates from the photometric redshift at the $\sim1.5\sigma$ level. \cite{Larson2018a} report a $2\sigma$ deviation of the photometric redshift from their $z=7.452$ galaxy, and photometric redshifts show a substantially higher fraction of catastrophic outliners ($\Delta z > 0.15$) in $z\sim$ 6 -- 7 \citep{Pentericci2018a}, compared to low-$z$ universe \citep{Dahlen2013a}.  Along with these recent findings, a statistical number of spectroscopic redshifts is required to precisely calibrate photometric redshifts in high-$z$ universe at $z>7$. 

The strong break between the optical and NIR bands implies that the emission line is either \lya (if it is the Lyman-break) or [\ion{O}{2}] $\lambda\lambda$3726, 3729 (the rest-frame 4000\AA/Balmer break). 
We note that other emission lines (e.g., H$\beta$ and [\ion{O}{3}] $\lambda\lambda$4959, 5007, or  H$\alpha$ with [\ion{N}{2}] $\lambda\lambda$6548, 6584) are additionally ruled out as the spectral coverage of our observations would allow us to detect multiple lines, and we do not see additional lines in the 1D or 2D spectrum.
In the case of the [\ion{O}{2}] doublet, the MOSFIRE spectral resolution can resolve the doublet with an expected gap of $\sim8-9$\AA\ (at $z=1.80$, the redshift if this line was [\ion{O}{2}]). Thus we analyzed the spectrum for signatures of the other line, with no signal observed at the expected wavelength. Furthermore, double Gaussian line fitting does not satisfy the theoretical expectation on the [\ion{O}{2}] doublet flux ratio of 0.35 -- 1.5 \citep{Pradhan2006a}. This low-$z$ solution is also disfavored by galaxy SED fitting with a much larger $\chi^2 (=23.4)$ than that of high-$z$ solution ($\chi^2=2.4$) (Figure 2; details described in Section 3.4). Finally, the best fit low-$z$ solution of the object suggests a very dusty but quenched galaxy with a near-zero star formation rate (SFR), inconsistent with strong [\ion{O}{2}] emission, which generally implies star formation (although it could also be caused due to active galactic nucleus activity).

\begin{table}[h]
\centering
\begin{center}
\caption{Summary of Emission Line Properties}
\begin{tabular}{lll}
\tableline 
\tableline
\quad{} & {z7\_GND\_16863} & {z7\_GND\_42912}\\
\tableline
\quad {$F_{\text{Ly}\alpha}$ (10$^{-17}$ erg s$^{-1}$ cm$^{-2}$)} & {$1.89\pm0.18$} & {$1.46\pm0.14$}\\
\quad {Signal-to-noise Ratio} & {$10.79$} & {$10.81$}\\
\quad {EW$_{\text{Ly}\alpha}$ (\AA)} & {$61.28\pm5.85$} & {$33.19\pm3.20$}\\
\quad {$z_{\text{Ly}\alpha}$} & {$7.599\pm0.001$} & {$7.506\pm0.001$}\\
\quad {$\sigma_{\text{blue}}$ (\AA)} & {$0.20^{+1.09}_{-0.20}$} & {$1.39^{+0.42}_{-0.35}$}\\
\quad {$\sigma_{\text{red}}$ (\AA)} & {$6.25^{+1.29}_{-1.05}$} & {$4.11^{+1.00}_{-0.91}$}\\
\quad {$\sigma_{\text{red}}/\sigma_{\text{blue}}$} & {$>4.29$\tablenotemark{a}} & {$2.98^{+1.64}_{-1.11}$}\\
\quad {FWHM$_{\text{red}}$ (\AA)\tablenotemark{b}} & {$14.71^{+3.03}_{-2.46}$} & {$9.68^{+2.36}_{-2.15}$}\\

\tableline
\end{tabular}
\end{center}
\begin{flushleft}
\tablenotetext{a}{1$\sigma$ lower limit}
\tablenotetext{b}{FWHM of the red side of the line (2.355$\sigma_{\text{red}}$)}
\end{flushleft}
\label{tab:LAEs}
\end{table}

\subsection{z7\_GND\_42912: a \lya emitter at $z=7.51$}
Our targets also include z7\_GND\_42912, shown in Figure 1(b), which was first reported in \cite{Finkelstein2013a} as a new \lya emission detection at $z=7.51$ with a S/N ratio of 7.8 from $\sim$6 hr of MOSFIRE observations in 2013 April.  Here we update the measure of the \lya line profile from the entire MOSFIRE dataset from three masks: GOODSN\_Mask1 (2013 Apr), GOODSN\_Y\_v12 (2014 Mar), and Mask2\_Y\_2015A (2015 Feb) with $t_{\text{exp}}$ = 16.5 hrs (bottom right in Figure 1b). With $\sim3\times$ longer exposure time, we reveal a clear asymmetric line profile with the updated line flux, $F_{\text{Ly}\alpha} =1.46\pm0.14 \times 10^{-17}$ erg s$^{-1}$ cm $^{-2}$ (S/N $=10.8$).

We note that our \lya flux measurement from z7\_GND\_42912 is $\sim$5$\times$ greater than that of \cite{Finkelstein2013a}. \cite{Tilvi2016a} published a \textit{HST}/grism observation of z7\_GND\_42912, finding a $\sim$4 times higher \lya flux than that of \cite{Finkelstein2013a}.  Although the origin of the significant discrepancy between HST/grism and Keck/MOSFIRE observations was not clearly known by the time, we found that this is mainly due to an unknown flux calibration issue in the \cite{Finkelstein2013a} analysis. The c.\ 2013 version of the MOSFIRE DRP gives different units between the multi-object spectra frames (electrons/sec) and longslit frames (ADU/coadd), but this difference in data units was not documented in the DRP documentation, thus these images were treated in the same manner in \cite{Finkelstein2013a}. By converting between these two image units, we find that the Finkelstein et al.\ (2013) line flux should be 4.65$\times$ higher, consistent with our updated measurement. Our updated flux with the same dataset (GOODSN\_Mask1) is $F_{\text{Ly}\alpha} =1.28\pm0.13 \times 10^{-17}$ erg s$^{-1}$ cm $^{-2}$, now consistent with that of \cite{Tilvi2016a}.  The flux values from the other individual masks are 1.48 and $1.01 \times 10^{-17}$ erg s$^{-1}$ cm $^{-2}$ with somewhat different line profiles (GOODSN\_Y\_v12 and Mask2\_Y\_2015A, respectively).  Thus, our final flux value in Table 2, which is measured from all combined data, is higher than the \cite{Tilvi2016a} value.  We note that while the significant variation in measured line flux between our three observations could imply a systematic uncertainty in our flux calibration, simulations have predicted that the measured Ly$\alpha$ flux can depend on the observed slit position angle due to the complicated morphology of the Ly$\alpha$ emission \citep{Smith2018a}.

\subsection{\lya emission properties}
The measured emission line properties of the two \lya emitting galaxies are listed in Table 3. From z7\_GND\_16863, the \lya emission line has been detected with $F_{\text{Ly}\alpha} =1.89\pm0.18 \times 10^{-17}$ erg s$^{-1}$ cm $^{-2}$ (S/N $=10.8$). Although the noise level is very high at the blue side of the line profile due to the nearby sky emission line, the asymmetric feature is still significant, with a 1$\sigma$ lower limit on $\sigma_{\text{red}}/\sigma_{\text{blue}}>4.29$.  Our new observations of the \lya emission line from z7\_GND\_42912 now reveal a significant asymmetric profile of $\sigma_{\text{red}}/\sigma_{\text{blue}}=2.98^{+1.64}_{-1.11}$.  This significant asymmetry was not found by Finkelstein et al. (2013), but is revealed in our higher fidelity spectrum. We also calculate the skewness of the \lya emission line profiles, which also suggests significant asymmetry on both emission lines with $1.37\pm0.23$ (z7\_GND\_16863) and $1.69\pm 0.20$ ( z7\_GND\_42912).  

Estimating \lya asymmetry, \cite{Rhoads2003a} introduced the parameters $a_{f}$ and $a_{\lambda}$ which represent the relative flux ratio between blue and red sides of \lya emission and the relative peak location from the blue and red ends of the line profile. \cite{Dawson2007a} performed a statistical study of $a_{f}$ and $a_{\lambda}$ with 59 \lya emitting galaxies, reporting significant asymmetric \lya emission profiles at $z\sim4$. We also estimate $a_{f}$ and $a_{\lambda}$ of our \lya emission lines with $a_{f}>3.70$ and $a_{\lambda}>3.43$ for z7\_GND\_16863 and $a_{f}>1.77$ and $a_{\lambda}>1.67$ ($1\sigma$ lower limits), further proving their asymmetry. Furthermore, more recent \lya surveys investigate \lya profiles, finding significant asymmetry of \lya emission lines at $4<z<7$ \citep{Ouchi2010a, Hu2010a, Kashikawa2011a, Mallery2012a, U2015a}. However, not many high-$z$ \lya emission lines have reported a significant asymmetric \lya line profile at $z>7$, presumably due to low signal-to-noise, though we note the stacked analysis of \lya emission shows a clear asymmetric line profile in \cite{Pentericci2018a}.  In addition to \cite{Song2016b}, which captured the first notable asymmetric line profile with deep NIR spectroscopy with 10 hrs of integration time, our analysis of \lya line profile with the extremely deep spectroscopy uncovers the asymmetric nature of our two \lya emission lines at $z>7$.

The asymmetric feature of \lya emission from high-$z$ galaxies is theoretically expected due to absorption by the interstellar medium (ISM) and IGM.  Interaction with an outflowing ISM provides easier escape routes for the red wing of \lya \citep[e.g.,][]{Ahn2001a,Dijkstra2014a}, thus the redshifted asymmetric \lya emission line profile is often explained by common galactic outflows \citep[e.g.,][]{Verhamme2006a, Gronke2015a, Remolina2019a}, which could be boosted by cosmic ray \citep{Gronke2018a} and \lya feedback \citep{Smith2017a,Kimm2018a}.  Importantly, recent studies on \lya profiles with Green Peas, a local analogue of a high-$z$ LAEs, have revealed more complex processes related to their \lya profiles \citep[e.g.,][]{Yang2016a, Yang2017a, Yang2017b, Verhamme2018a, Orlitova2018a}. Therefore, further studies on \lya profiles are yet required to illustrate the detailed \lya radiative processes in ISM.

 To place \lya detection limits for our non-detections, we calculate 5$\sigma$ detection limits of \lya emission lines in the MOSFIRE Y-band wavelength coverage from $\sim9800$ -- $11200$\AA\ by adding mock 1D \lya emission lines on the actual 1D spectra.  For simulating the mock \lya emission lines, we renormalize the best fit asymmetric Gaussian profile from the detected \lya emission in z7\_GND\_42912.  The measured median 5$\sigma$ detection limit of \lya emission is down to $\sim4\times10^{-18}$ erg s$^{-1}$ cm $^{-2}$ between sky lines. This measurement is consistent with the previous observations \citep[e.g.,][]{Wirth2015a, Song2016b} when scaling our detection limit by $\sqrt{t}$, where $t$ is the integration time. Compared to the emission line sensitivity in the MOSDEF survey \citep{Kriek2015a}, our \lya sensitivity is lower by up to a factor of two (for an optimistic case of MOSDEF). The difference is understandable as MOSDEF has overall better seeing than our data, and the \lya line profile is generally broader than other emission lines, making it more difficult to detect. 
 
 We measure the rest-frame EWs of the detected \lya lines and place $5\sigma$ EW upper limits for non-detections. The rest-frame EW is defined as the ratio of the \lya flux to the UV continuum flux density, divided by $1+z$. We derive the UV continuum brightness from the best-fit galaxy spectral energy distribution (SED) model (refer to Section 3.4), averaged over the 50\AA\ window just redward of \lya emission, from 1230\AA -- 1280\AA.  The derived EWs are $61.28\pm5.85$\AA\ and $33.19\pm3.20$\AA\ for z7\_GND\_16863 and z7\_GND\_42912, respectively.  Previous measures of \lya EWs in the literature show a deficit of high EW LAEs ($>$50\AA) at $z>7$ \citep[e.g.,][and references therein]{Tilvi2014a}, and the measured EW of the \lya emission line from z7\_GND\_42912 ($33.19\pm3.20$\AA) is consistent with those measurements. However, the \lya EW from the new \lya emission line in z7\_GND\_16863 ($61.28\pm5.85$\AA) is relatively high. Along with the recent observations of \cite{Hu2017a}, \cite{Zheng2017a}, and \cite{Pentericci2018a} which found high EW LAEs at $z\sim7$ and \cite{Larson2018a} which reported a high EW object ($140.3\pm19.0$\AA) at $z\sim7.5$, our results show that high-EW Ly$\alpha$ emission is not uncommon at $z >$ 7.
 
 Even with our very deep NIR observations, we do not detect \lya emission from z7\_GND\_18869 and z7\_GND\_9408. However, those are faint objects, having large EW upper limits ($<$103 and $<$387 \AA, respectively), thus non detections from the two faint galaxies are well expected and understandable.  However, detecting \lya lines from two out of two observed bright sources is somewhat unexpected, due to previous results implying an increasing neutral fraction of the IGM at $z>7$.

The MOSFIRE $Y$-band covers the wavelength ranges of \ion{N}{5} emission lines, an indicator of active galactic nuclei (AGN) activity, for our two LAEs.  For z7\_GND\_16863, we search within 500 km s$^{-1}$ from the expected wavelength \citep[e.g.,][]{Steidel2010a, Erb2014a, Stark2017a, Mainali2018a}, finding no significant detection, and a 1$\sigma$ upper limit on the \ion{N}{5} emission line flux of $\lesssim$8.54$\times10^{-19}$ erg s$^{-1}$ cm $^{-2}$, corresponding to the \lya/\ion{N}{5} flux ratio $\gtrsim$22.  For z7\_GND\_42912, Hutchison et al. (2019, in prep) detect one of the \ion{C}{3}] lines with MOSFIRE $H$-band observations, measuring the systematic redshift to be $z=7.5027\pm0.0003$ and $7.4941\pm0.0003$, if their detected line is \ion{C}{3}] $\lambda$1907 and \ion{C}{3}] $\lambda$1909, respectively.  We do not see significant emission for \ion{N}{5} at the wavelengths corresponding to these systemic redshifts, and we measure 1$\sigma$ upper limits of $<$7.66 and $<$6.60$\times10^{-19}$ erg s$^{-1}$ cm$^{-2}$ at 10543 and 10532\AA, corresponding to \lya/\ion{N}{5} flux ratios of $>$19 and $>$22, respectively.  \cite{Tilvi2016a} measured a possible detection of \ion{N}{5} from z7\_GND\_42912 at $\lambda\sim10550$\AA\ with a slight spatial offset of $\sim0\farcs1$ from \textit{HST}/grism observations, with a reported \ion{N}{5} line flux of $f_{\text{line}}=0.91\pm0.21\times10^{-17}$ erg s$^{-1}$ cm $^{-2}$.  Our spectrum should have detected this line with S/N $>$ 10.  This our non-detection implies that the previously reported \ion{N}{5} in \cite{Tilvi2016a} may be contamination.  The previously reported measures of the \lya/\ion{N}{5} flux ratio from several $z\gtrsim7$ galaxies range from $\sim$1 -- 2 \citep{Hu2017a, Sobral2017a} to $\sim$6 -- 9 \citep{Laporte2017a, Mainali2018a}.  With the limits of the \lya/\ion{N}{5} flux ratio $\gtrsim$19 -- 22 in our observations, our two LAEs likely do not host significant AGN activity \citep[see also discussion in][]{Castellano2018a}.

\begin{figure}[t]
\centering
\gridline{\fig{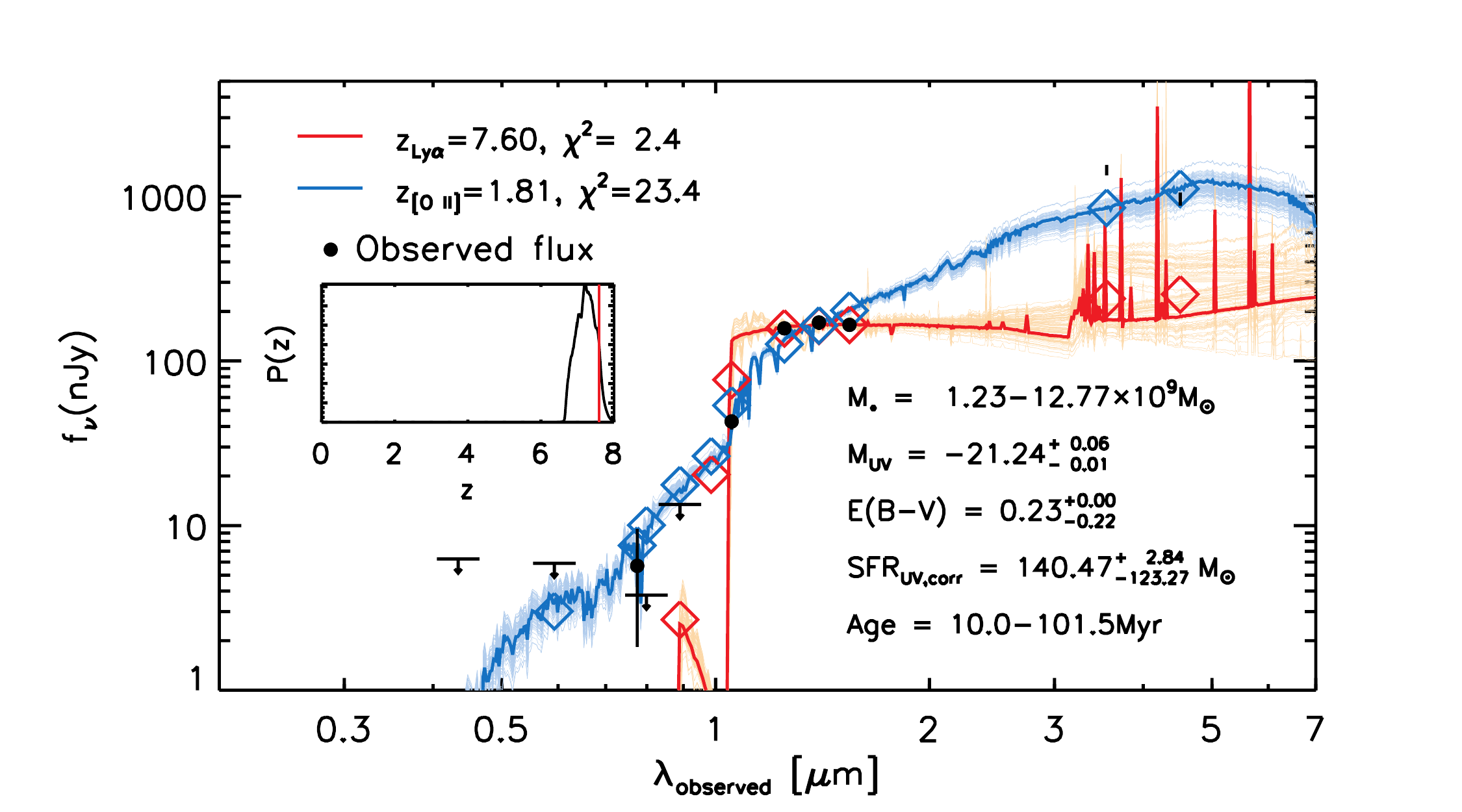}{1.1\columnwidth}{(a) z7\_GND\_16863}}
\gridline{\fig{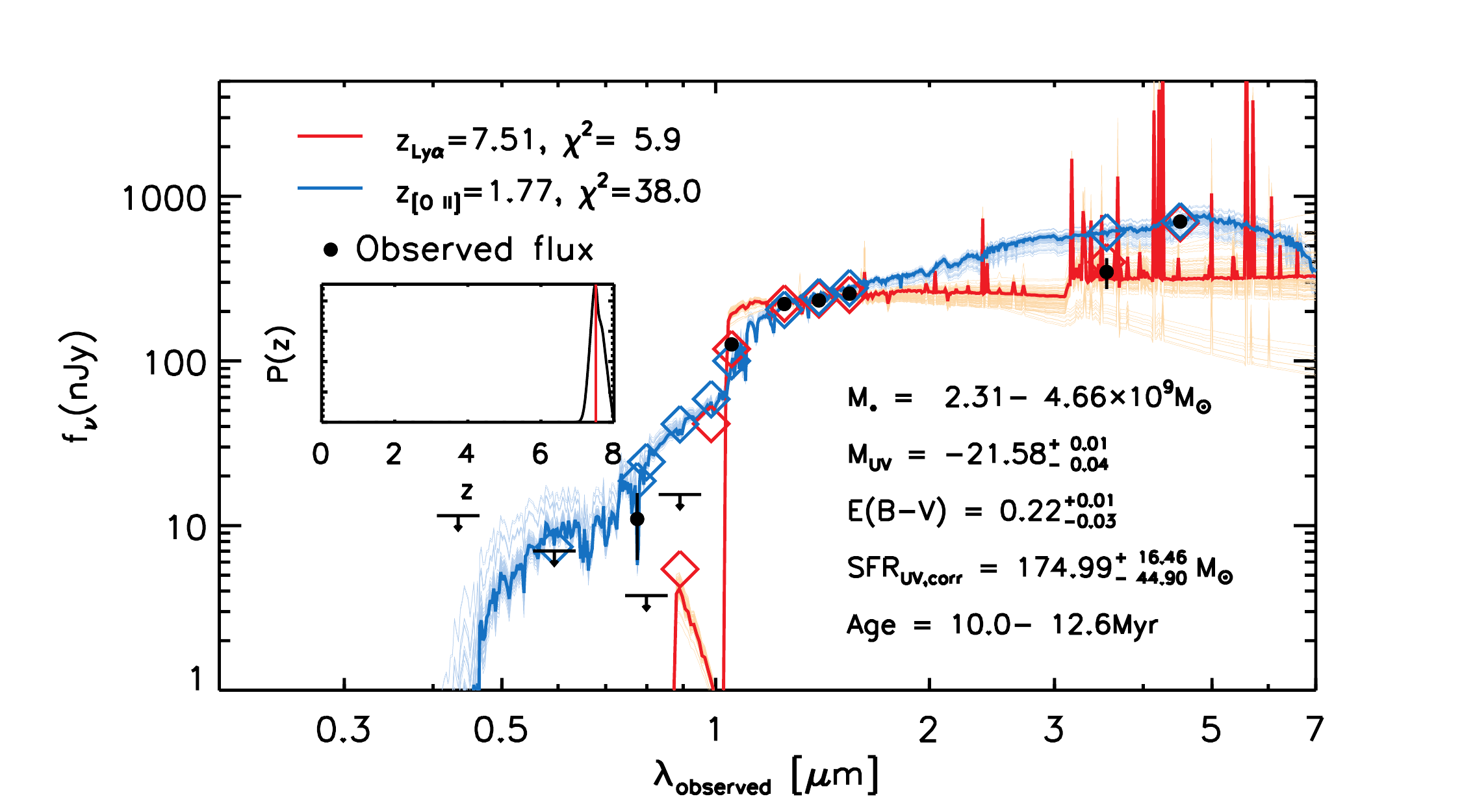}{1.1\columnwidth}{(b) z7\_GND\_42912}}
\caption{Galaxy SED fitting results. Each panel shows two SEDs for high-$z$ (Ly$\alpha$) and low-$z$ ([\ion{O}{2}]) solutions (red and blue solid curves, respectively), and colored diamond symbols represent continuum fluxes from the model SEDs.  Black dots are observed fluxes with their associated errors, and the downward arrows are 1$\sigma$ upper limits. The physical quantities written in the panels are estimated from the high-$z$ solutions. For stellar masses and ages, we display 68\% confidence ranges. Insets display photometric redshift probability distributions, P(z), taken from Finkelstein et al. (2019, in prep), and the spectroscopic redshifts are shown with vertical red lines.}
\label{fig:sed}
\end{figure}

\subsection{Physical Properties of $z\sim7.5$ Galaxies}
To derive the physical quantities of our detected galaxies, we perform galaxy spectral energy distribution (SED) fitting with stellar population synthesis models \citep{Bruzual2003a}. The parameter settings, nebular emission lines \citep{Inoue2011a, Salmon2015a}, and IGM attenuation \citep{Madau1995a} descriptions in SED fitting are similar to those used in \cite{Jung2017a}. We take a \cite{Salpeter1955a} initial mass function with lower and upper stellar-mass limits of 0.1 to 100 M$_{\odot}$, respectively, and metallicities range from 0.005 to 1.0 $Z_{\odot}$. We allow several star-formation histories (SFHs), using a range of exponential models: decrease, increase, and constant SFHs with exponential-decaying time ($\tau =$10Myr, 100Myr, 1Gyr, 10Gyr, 100Gyr, -300Gyr, -1Gyr, -10Gyr). Dust attenuation to our model spectra uses the attenuation curve of \cite{Calzetti2001a} with $E(B-V)$ values spanning 0 -- 0.8.  We restrict stellar population ages to be $>$ 10 Myr, to avoid a scenario where a galaxy forms all of its mass in an unphysically small amount of time.

Constraining the stellar mass for high-$z$ galaxies is critically dependent on long wavelength $Spitzer/$IRAC photometry.  However, z7\_GND\_16863 is found in the vicinity of a bright nearby source (Figure1a), making it impossible to properly measure its rest-frame optical continuum with the low spatial resolution IRAC photometry.  Thus, our SED fitting results of z7\_GND\_16863 with only rest-frame UV fluxes have highly correlated physical parameters (e.g., stellar mass, dust extinction, and age). The \lya contributions to continuum fluxes were removed during the SED fitting, and we also ignored $Y_{105}$ fluxes as it is often difficult to model due to the uncertainty of IGM attenuation. To calculate SFRs, we first obtained dust-corrected UV fluxes from the best fit models and convert the UV fluxes to SFRs by adopting the updated FUV-to-SFR conversion factor of $\kappa = 1.15 \times 10^{-28} M_{\odot}$ yr$^{-1}$ erg$^{-1}$ s Hz (Madau \& Dickinson 2014). This is derived from the stellar population models of \cite{Conroy2009a} and $\sim20$\% lower than the conventional conversion factor in \cite{Kennicutt1998a}. This is also similar to the other recent studies \citep[e.g.,][]{Salim2007a, Haardt2012a} based on \cite{Bruzual2003a} which found lower mean conversion factors.

Figure 2 shows our SED fitting results for z7\_GND\_16863 (top) and z7\_GND\_42912 (bottom).  The best fit model is chosen by minimizing $\chi^2$. The fitting errors are obtained via Monte-Carlo (MC) simulations. We run 1000 MC simulations to derive the best fit models with the simulated continuum fluxes. In the MC simulations, we fluctuate the observed fluxes with Gaussian random deviates which is equivalent to the flux measurement errors to simulate the continuum fluxes, and we perform SED fitting with the simulated fluxes.  

z7\_GND\_16863 and z7\_GND\_42912 are very bright in the rest-frame UV ($M_{\text{UV}} = -21.24$ and $-21.58$, respectively), relative to the characteristic UV magnitudes of $z\sim7-8$ galaxies of $M_{\text{UV}} \sim 21$ (Finkelstein et al. 2015).  Stellar masses of the two objects are consistent with that expected from published scaling relation between M$_{*}$ -- $M_{\text{UV}}$ at $z>7$ \citep{Song2016a}, although the stellar mass of z7\_GND\_16863 is not well-constrained due to the lack of the rest-frame optical photometric constraints.  The two \lya emitters are actively forming stars with SFRs [M$_{\odot}$ yr$^{-1}$] = $140^{+3}_{-123}$ (z7\_GND\_16863) and  $175^{+16}_{-45}$ (z7\_GND\_42912).  These SFRs are above the fiducial M$_{*}$ -- SFR relation of high-$z$ galaxies \citep[e.g.,][]{Salmon2015a, Song2016b, Jung2017a} which suggests $\sim$10 -- 30 M$_{\odot}$ yr$^{-1}$ for $\sim10^{9-10}$M$_{\odot}$ galaxies at $z\sim6$, although this relation likely increases at higher redshift. 
Particularly, our updated measurement of z7\_GND\_42912 with young stellar populations (age $<12.6$ Myr) and a high SFR is yet comparable but less extreme than \cite{Finkelstein2013a}, which found the time-averaged SFR of z7\_GND\_42912 higher than 330 M$_{\odot}$ yr$^{-1}$ with extremely young stellar populations (age $\sim$1--3 Myr [$1\sigma$], which is less than we allowed in our model fitting). We also calculate time averaged SFRs by simply dividing the stellar mass by the stellar population age. The averaged SFRs of the galaxies are $\gtrsim$85 and $\gtrsim$216 M$_{\odot}$ yr$^{-1}$ in their $1\sigma$ low limits for z7\_GND\_16863 and z7\_GND\_42912, respectively.
Even with their large uncertainties, our $z\sim7.5$ \lya emitting galaxies require high SFRs to build up their stellar masses.  Such high SFRs are expected for producing their asymmetric \lya profiles with strong galactic outflows due to stellar feedback.

\section{Summary and Discussion}
We analyze our deepest NIR spectroscopic observations with Keck/MOSFIRE for four target galaxies at $z_{\text{phot}}\gtrsim7$ with $\gtrsim16$ hr of integration time. We detect two \lya emission lines from UV-bright and actively star-forming galaxies, discovering a new \lya emitting system at $z=7.60$ (z7\_GND\_16863) as well as providing an updated measure of a $z=7.51$ \lya emission line (z7\_GND\_42912) which was previously reported in \cite{Finkelstein2013a} and \cite{Tilvi2016a}.  We measure the detailed \lya line profiles, finding significant \lya asymmetry.  The two detected \lya emission lines from bright sources ($M_{\text{UV}}<-20.25$) could imply that these bright galaxies likely inhabit ionized bubbles in a partially neutral IGM, although deeper exposures may yet reveal \lya emission in the fainter sources.

With the current consensus from \lya studies around the end of reionization at $z\sim6-7$, \lya visibility is expected to decrease as the IGM becomes neutral into the epoch of reionization, and it is also expected to have a significant dependence on the UV brightness of galaxies. Conventionally, galaxies are divided into bright ($M_{\text{UV}}<-20.25$) and faint ($M_{\text{UV}}>-20.25$) groups, and the \lya fraction is observationally suggested to be higher from faint sources than that from bright ones \citep[e.g.,][and references therein]{Stark2016a}.  Particularly, \cite{Pentericci2018a} found a higher \lya fraction among faint samples again at $z\sim7$.  This is explained by bright galaxies being more likely to be evolved, with a higher metallicity and larger amount of dust, reducing the \lya photon escape probability. 

On the contrary, previous results at $z>7$ \citep{Stark2017a} suggest that very bright galaxies reside in ionized bubbles, allowing a larger transmission of \lya emission than faint sources. Similarly, \cite{Castellano2018a} found more \lya detections than expected among bright galaxies, while they failed to find faint galaxies which emit \lya photons at $z\sim7$. A possible explanation is that brighter galaxies reside in early overdensities, which are ionized earlier compared to the rest of the Universe.  In addition, as discussed in \cite{Mason2018a}, \lya photons from bright sources could escape easier due to their higher velocity offsets from systemic, making them less affected by neutral hydrogen in the IGM.

Furthermore, \cite{Zheng2017a} reports a "bump" at the bright end of the \lya LF at $z\sim7$ from the Lyman Alpha Galaxies in the Epoch of Reionization (LAGER) survey, indicative of large ionized bubbles where we could see different evolution at bright and faint ends of the \lya LF.  In the same context, our two detected \lya emission lines from bright sources are suggestive that the \lya visibility of UV-bright galaxies does not decrease as much as that of faint galaxies at $z > 7$.  Of course, this is highly tentative as the number of targets in our study is  small, and the \lya EW detection limits for our faint sources are not as deep as those for the bright ones.  Therefore, a comprehensive analysis is necessary to assure if the \lya detection rate is higher among brighter galaxies at $z>7$.

Our entire MOSFIRE dataset will be included in the next publication where we will place a strong constraint on the \lya visibility into $z>7$ with the most comprehensive dataset of NIR spectroscopic follow-up observations. Furthermore, \lya studies in the even earlier universe is promising with the \textit{James Webb Space Telescope} NIR spectroscopy \citep{Smith2018a}, and a future \lya survey project with the extremely large telescopes (e.g., the Giant Magellan Telescope) will deliver an extensive \lya dataset, which allows us to explore large areas and study the topology of reionization.

\acknowledgments
The authors wish to acknowledge the very significant cultural role and reverence that the summit of Mauna Kea has always had within the indigenous Hawaiian community.  We are most fortunate to have the opportunity to conduct our observations from this mountain.  I.J. acknowledges support from the NASA Headquarters under the NASA Earth and Space Science Fellowship Program - Grant 80NSSC17K0532.  I.J. and S.F. acknowledge support from NSF AAG award AST-1518183.

\listofchanges
\end{document}